# An Efficient Method for Evaluating the Feasibility of Spaceborne SAR for Ocean Ship Detection

Anatolii A. Kononov, *Independent Researcher*

*Abstract*— This letter presents an effective method for assessing the feasibility of detecting ocean ships using spaceborne synthetic aperture radar (SAR). The technique employs the minimum detectable radar cross-section criterion under specified false alarm and detection probabilities. The benefits of the proposed method are illustrated by evaluating the feasibility of detecting small ships with SAR from a satellite in very low Earth orbit.

## I. INTRODUCTION

MONITORING and detecting illegal fishing vessels and small boats within a nation's exclusive economic zones is crucial for financial protection and national security. However, traditional methods, such as patrol ships and sporadic aerial surveillance, often prove ineffective due to high costs and limited coverage, allowing many unauthorized vessels to evade detection.

Spaceborne Synthetic Aperture Radar (SAR) operating in very low Earth orbit (VLEO) satellites (Fig. 1) presents a promising alternative by offering consistent, weather-independent, day-and-night ocean surveillance. A constellation of small satellites is necessary due to the limited ground coverage of typical SAR systems and the need for near-real-time monitoring. Each satellite must have an SAR sensor optimized for detection performance and cost, making early-stage feasibility assessments critical.

A preliminary feasibility assessment method was recently introduced in [1]. This method employs a simplified two-step detection algorithm that assumes receiver thermal noise is the only source of interference. The first step employs a constant false alarm rate (CFAR) detector to identify SAR image pixels that exceed a predefined threshold, corresponding to a specified pixel-level probability of false alarm, $P_{fa}$. The second step uses a target detection window (TDW) to convolve it with the CFAR pixel map. This step employs a binary integration scheme ($m$-of-$n$ decision rule) in which a ship is declared present if at least $m$ pixels exceed the threshold within the TDW.

However, the method [1] has fundamental limitations. First, it fails to establish a solid criterion for assessing the feasibility of ship detection. Second, it depends on time-consuming Monte Carlo simulations to estimate the overall detection probability, $P_D$. Third, it employs an unspecified approximation procedure to calculate $P_D$ and the ship-level false alarm probability, $P_{FA}$. This procedure computes binomial coefficients, which become numerically unmanageable for large $n$. Fourth, the method lacks a clearly defined criterion for selecting the optimal $m$ value that maximizes $P_D$.

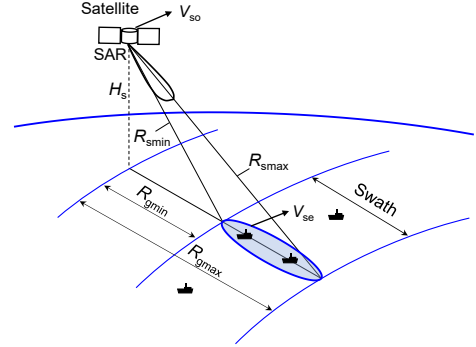

Fig. 1.  SAR for ocean ship detection from a VLEO satellite

This letter proposes an improved approach to analyzing the feasibility of SAR-based ship detection. We establish a solid criterion for assessing ship detection feasibility. We derive a straightforward formula for pixel-level detection probability, $P_d$, thereby eliminating the need for Monte Carlo simulation. Additionally, we introduce an efficient and numerically stable method for computing ship-level probabilities $P_D$ ($P_{FA}$) based on pixel-level probabilities $P_d$ ($P_{fa}$), and vice versa. Finally, we establish a rigorous rule for selecting the optimal value of $m$ that maximizes the ship-level detection probability $P_D$.

We demonstrate the advantages of the proposed technique by evaluating the feasibility of ship detection with an SAR in a very low-Earth orbit satellite.

## II. SAR IMAGE DATA MODEL

This section uses a SAR image data model from [1]. The model assumes that the only interference is the receiver's thermal noise of known power. Consequently, the pixel data in the SAR image may represent either the intensity of the background noise alone or the returns associated with ships.

The statistical analysis conducted in [1] for TanDEM-X data indicates that the backscattering coefficient distribution for the ship-associated pixels is well-approximated by a lognormal distribution with the probability density function (PDF)

$$p_{\Sigma_{sp}^0}(\sigma) = \frac{1}{\sigma\beta\sqrt{2\pi}} e^{-\frac{(\ln\sigma-\alpha)^2}{2\beta^2}}, \sigma \geq 0 \quad (1)$$

In (1) above, the symbol $\Sigma_{sp}^0$ denotes the random backscattering coefficient, while the distribution parameters $\alpha$ and $\beta$ represent the mean and standard deviation of the associated Gaussian random variable $Y = \ln\Sigma_{sp}^0$: $Y \sim \mathcal{N}(\alpha, \beta^2)$ and $\Sigma_{sp}^0 \sim \mathcal{LN}(\alpha, \beta)$.

For model (1), given $\alpha$ and $\beta$, the mean value of the ship's backscattering coefficient is

$$\bar{\Sigma}_{sp}^0 = e^{\alpha+\beta^2/2} \quad (2)$$

Work [1] provides a set of estimates for $\alpha$ and $\beta$ calculated for each target as a function of ship size in pixels using data from TanDEM-X images that include 58 targets of varying size.

In the present study, we assume the parameter $\beta = 2$; this value is close to estimates for ships with a size $N_{sp} \leq 600$ ([1], Fig. 4). Given $\bar{\Sigma}_{sp}^0$ and the parameter $\beta$, one can calculate the parameter $\alpha$ as

$$\alpha = \ln\bar{\Sigma}_{sp}^0 - \beta^2/2 \quad (3)$$

To evaluate the minimum detectable RCS of a ship one needs to calculate the average signal-to-noise ratio ($\bar{X}_{sp}$) for the pixels associated with the ship. This quantity is provided by [2]

$$\bar{X}_{sp} = \frac{P_{avg} G^2 \lambda^3 \bar{\Sigma}_{sp}^0 \delta_r}{2(4\pi)^3 R^3 \, kT_0 F \, L_s \, V_{so} \cos\psi} \quad (4)$$

where $P_{avg}$ is the average transmit power, $G$ is the antenna gain, $\lambda$ the radar wavelength, $R$ is the target range, $k = 1.380 \cdot 10^{-23}$J/K is the Boltzmann's constant, $T_0 = 290K$ is the standard temperature, $F$ is the noise factor, $L_s$ is the system loss, $V_{so}$ is the satellite orbital speed, and $\psi$ is the grazing angle.

The parameter $\delta_r$ represents the pixel width in slant range; its value is $\delta_r = K_{pw} c/(2B)$, where $c = 3 \cdot 10^8$ m/s is the speed of light, $B$ is the transmit waveform bandwidth, and $K_{pw}$ is the pulse widening constant. In this study, we assume $K_{pw} = 1.5$.

From (4), the randomly fluctuating signal-to-noise ratio (SNR) $X_{sp}$ for the ship-related pixel is directly proportional to $\Sigma_{sp}^0$; specifically, $X_{sp} = a\Sigma_{sp}^0$, where the constant $a$ defined as

$$a = \frac{P_{avg} G^2 \lambda^3 \delta_r}{2(4\pi)^3 R^3 \, kT_0 F \, L_s \, V_{so} \cos\psi} \quad (5)$$

Since $X_{sp}$ is a linear transformation of the random variable $\Sigma_{sp}^0$, the PDF of $X_{sp}$ in terms of the PDF of $\Sigma_{sp}^0$ is expressed as [3]

$$p_{X_{sp}}(\chi) = \frac{1}{a} p_{\Sigma_{sp}^0}\left(\frac{\chi}{a}\right), \chi \geq 0 \quad (6)$$

Applying (1) and (6) yields

$$p_{X_{sp}}(\chi) = \frac{1}{\chi\beta\sqrt{2\pi}} e^{-\frac{[\ln\chi-\alpha']^2}{2\beta^2}}, \chi \geq 0 \quad (7)$$

As follows from (7), $X_{sp} \sim \mathcal{LN}(\alpha', \beta)$, where $\alpha' = \alpha + \ln a$. Therefore, the SNR for the ship-associated pixel also follows a lognormal distribution with the same beta-parameter $\beta$ as $\Sigma_{sp}^0$, but a different alpha-parameter $\alpha'$. As a result, the mean value of the SNR for the ship-associated pixel is

$$\bar{X}_{sp} = e^{\alpha'+\beta^2/2} \quad (8)$$

Given the average SNR $\bar{X}_{sp}$ and parameter $\beta$, one can calculate the new alpha-parameter from (8) as

$$\alpha' = \ln\bar{X}_{sp} - \beta^2/2 \quad (9)$$

III. DETECTION ALGORITHM

This section outlines a two-step algorithm for assessing the feasibility of ship detection using spaceborne SAR. The first step employs a constant false alarm (CFAR) global threshold at the individual pixel level across the SAR image. The second step performs final target detection for the remaining pixels by applying the principle of binary integration. This method counts the number of pixels within target detection windows that cover the entire SAR image and compares these counts to the minimum integer threshold needed to establish detection.

Each pixel in a SAR image represents the intensity of thermal receiver noise or ship-related intensity governed by a lognormal distribution introduced in section II. Therefore, the complex sample of an individual pixel can be expressed as

$$U = \begin{cases} \sqrt{X_{sp}}e^{j\Phi} + \mathcal{N}: & \text{if target present} \\ \mathcal{N} & : \text{if no target} \end{cases} \quad (10)$$

where $X_{sp} \sim \mathcal{LN}(\alpha', \beta)$ is the intensity of the single ship-related pixel, $\Phi \sim \mathcal{U}(0,2\pi)$ is the phase uniformly distributed over the interval $[0, 2\pi]$, and $\mathcal{N} \sim \mathcal{CN}(0,1)$ is the circularly symmetric complex Gaussian noise of unit power.

For model (10), the global thresholding is given by

$$\begin{cases} \text{if } Z \geq T: \text{sample } z \text{ survives} \\ \text{if } Z < T: \text{sample } z \text{ nullified} \end{cases} \quad (11)$$

where $Z = |U|^2$ represents the intensity of the pixel's complex value, $T = -\ln P_{fa}$ is the detection threshold, where $P_{fa}$ is the probability of false alarm at the individual pixel level. Although the threshold $T$ is constant, procedure (11) exhibits the CFAR property because the noise power is assumed to be known.

The required $P_{fa}$ value is determined by the overall probability of false alarm $P_{FA}$ for the area of interest (AOI). Let the area of an AOI be $A_{AOI}$, and the area of a resolution cell be $A_{res} = \delta_{az}\delta_{gr}$, where $\delta_{az}$ is the azimuth (or cross-range) resolution, and $\delta_{gr} = \delta_r/\cos\psi$ is the ground range resolution, respectively.

The total number of pixels in the AOI is $N_{AOI} = \lceil A_{AOI}/A_{res} \rceil$, where $\lceil x \rceil$ is the smallest integer greater than or equal to $x$. Therefore, the required $P_{fa}$ value is [4, p.339]

$$P_{fa} = P_{FA}/N_{AOI} \quad (12)$$

For instance, if $P_{FA} = 10^{-5}$, $A_{AOI} = 25 \times 400$ km$^2$ = $10^{10}$ m$^2$, $\delta_{az} = 2$ m, $\delta_{gr} = 5$ m, then $A_{res} = 10$ m$^2$, $N_{AOI} = 10^9$, and from (12) we derive $P_{fa} = 10^{-14}$.

The second step involves a target detection window (TDW), which is a square with a side length of $L_w$ adjusted to the smallest expected target length, defined as $L_w = \min \{ L_{\text{ship}} \}$, where $\{ L_{\text{ship}} \}$ is a set of the lengths of the ships of interest.

Let $\delta_{\min}$ be defined as $\delta_{\min} = \min \{\delta_{\text{az}}, \delta_{\text{gr}}\}$. We define the side of the TDW in pixels as

$$P_w = \lceil L_w/\delta_{\min} \rceil \quad (13)$$

The total number of pixels in the TDW is $N_{\text{pw}} = P_w \times P_w$. This configuration of the TDW ensures the maximum number of ship-associated pixels within it, regardless of whether the ship's axis is parallel to the $x$ or $y$ coordinate axis.

For a ship of dimensions $L_{\text{ship}}$-by-$W_{\text{ship}}$, the upper bound for the ship's area is $A_{\text{ship}} = L_{\text{ship}} W_{\text{ship}}$. Therefore, the maximum number of ship-associated pixels that may appear in a SAR image is $N_{\text{ps}} = \lfloor A_{\text{ship}}/A_{\text{res}} \rfloor$.

The second step counts the number of surviving pixels in non-overlapping TDWs that compactly tiled over the azimuth-range grid associated with a SAR image to detect the ship-associated pixels. This final detection step employs the $m$-of-$N_{\text{ps}}^w$ rule

$$\begin{cases} \text{if } m_{ij}^{\text{ps}} \geq m: \text{ target is in TDW}_{ij} \\ \text{if } m_{ij}^{\text{ps}} < m: \text{ no target in TDW}_{ij} \end{cases} \quad (14)$$

where $\text{TDW}_{ij}$ denotes the $(i,j)$-th TDW that may contain a maximum of $N_{\text{ps}}^w$ ship-associated pixels (for a single ship), $m_{ij}^{\text{ps}}$ is the number of surviving pixels in the $(i,j)$-th TDW, and $m$ is the minimum number of surviving pixels that must be present to declare the ship's detection in the $(i,j)$-th TDW.

## IV. DETECTION PERFORMANCE

This section derives equations to evaluate the performance of the two-step detection algorithm described in Section III.

First, consider the step of global thresholding. As can be seen, equations in (10) present the "received signal model" in the detection problem of a signal with random amplitude ($\sqrt{X_{\text{sp}}}$) and phase ($\Phi$) amid white Gaussian noise, where both the signal and noise are represented by a single sample.

To derive the equation for the probability of detection $P_d$ for the detector (11) we employ the Bayesian approach. Assuming the amplitude in (10) is fixed allows us to easily obtain the conditional probability of detection $P_d(\chi), X_{\text{sp}} = \chi$, meaning $\chi$ is a fixed SNR value. Indeed, under this condition, (10) describes the detection of a non-fluctuating signal with random phase embedded in the white Gaussian noise. Under this condition, the probability of detection is given by [4, p. 316]

$$P_d(\chi) = Q_1\left(\sqrt{2\chi}, \sqrt{-2\ln P_{\text{fa}}}\right) \quad (15)$$

where $Q_1(\cdot,\cdot)$ is the Marcum's $Q$ function of order 1.

According to the Bayesian approach, we obtain the unconditional probability of detection $P_d$ by averaging $P_d(\chi)$ in equation (15) as shown below

$$P_d = \int_0^\infty Q_1\left(\sqrt{2\chi}, \sqrt{-2\ln P_{\text{fa}}}\right) p_{X_{\text{sp}}}(\chi) d\chi \quad (16)$$

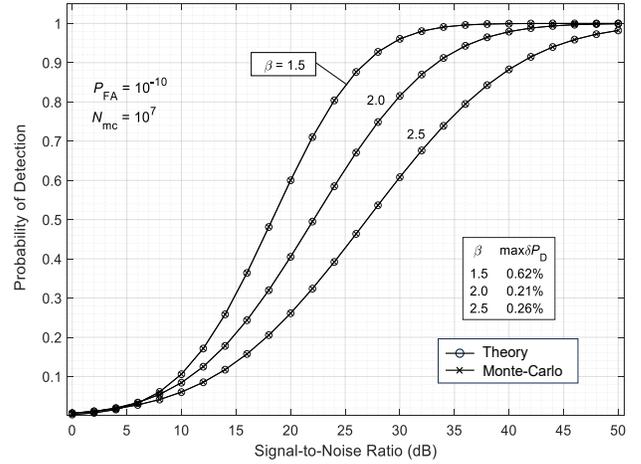

Fig. 2. Comparison between the theoretical $P_d$-vs-$\bar{X}_{\text{sp}}$ plots and the empirical $\hat{P}_d$-vs-$\bar{X}_{\text{sp}}$ plots at different $\beta$

In (16) above, the PDF $p_{X_{\text{sp}}}(\chi)$ is given by (7) with parameters $\alpha'$ and $\beta$, where $\alpha' = \ln \bar{X}_{\text{sp}} - \beta^2/2$, and the average SNR $\bar{X}_{\text{sp}}$ can be derived from (4) for any specified SAR parameters.

Thus, the probability of detection $P_d$ at the individual pixel level (during the global thresholding step) can be calculated efficiently from (16) using numerical integration. This method ensures highly accurate results and requires significantly less computational time than Monte Carlo simulations.

Fig. 2 validates the significance of (16). It contrasts the theoretical $P_d$-vs-$\bar{X}_{\text{sp}}$ plots, calculated through numerical integration in (16), with the empirical $\hat{P}_d$-vs-$\bar{X}_{\text{sp}}$ curves, where $\hat{P}_d$ is evaluated using $N_{\text{mc}} = 10^7$ Monte-Carlos. These plots are created at $P_d = 10^{-10}$ for $\beta$ values of 1.5, 2, and 2.5. For each $\beta$ value, the parameter $\alpha'$ is derived from (9) for the corresponding $\bar{X}_{\text{sp}}$ value. The maximum relative error $\max \delta P_d$, where $\delta P_d = 100|P_d - \hat{P}_d|/P_d$, remains below 1% across the range of 0 to 50 dB for the average SNR $\bar{X}_{\text{sp}}$.

Now, consider the overall detection performance of the algorithm in question. The overall probabilities of false alarm, $P_{\text{FA}}$, and detection, $P_D$, are simple to calculate. The final decision is based on the $m$-of-$n$ detection rule with $n = N_{\text{ps}}^w$ in the second step. Thus, given the probability of false alarm $P_{\text{fa}}$ and detection $P_d$ at the individual pixel level, the overall $P_{\text{FA}}$ and $P_D$ values at the ship level are given by [5]

$$P_F^{\text{sw}} = I_{P_f}(m, n - m + 1) \quad (17)$$

where $I_x(m, n - m + 1)$ is the regularized incomplete beta function, and a pair of symbols F and f represent the following combinations: FA and fa, to calculate $P_{\text{FA}}$ given $P_{\text{fa}}$, or D and d to determine $P_D$ given $P_d$.

Using (17), one can accurately and efficiently calculate the desired probabilities with built-in MATLAB functions or other software packages. Unlike standard equation for binary integration, this equation does not encounter numerical issues when $n$ is a large integer, such as that in [4], which fail when computing binomial coefficients for large $n$. Additionally, based on the $P_{FA}$ or $P_D$, the inverse $I_P^{-1}(m, n-m+1)$ of the regularized incomplete beta function allows accurate and efficient calculation of the $P_{fa}$ or $P_d$ value as [5]

$$P_f = I_{P_F}^{-1}(m, n-m+1) \tag{18}$$

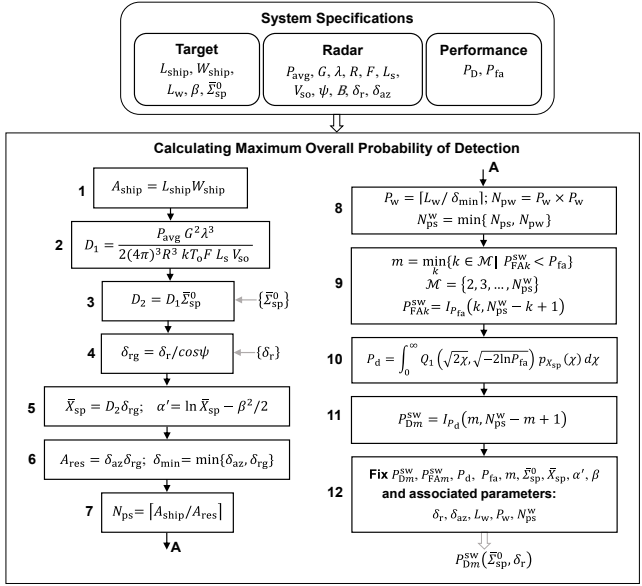

Fig. 3. Calculating maximum overall probability of detection

Fig. 3 illustrates a flowchart detailing the procedure for maximizing overall detection probability based on a set of fixed parameters obtained from specific system specifications. Blocks 1 and 2 calculate the ship area $A_{ship}$ and the auxiliary parameter $D_1$ to determine the pixel-associated average SNR $\bar{X}_{sp}$. Next, block 3 computes a $\bar{\Sigma}_{sp}^0$ dependent parameter $D_2$, while block 4 determines the pixel's ground extent in range $\delta_{gr}$ for the specified slant range resolution $\delta_r$ and grazing angle $\psi$. Block 5 calculates the pixel's average SNR value $\bar{X}_{sp}$ and the parameter $\alpha'$ given the parameter $\beta$ (a typical value for the ships of interest) for the pixel-associated PDF $p_{X_{sp}}(\chi)$, where $X_{sp} \sim \mathcal{LN}(\alpha', \beta)$ as provided by (7).

Blocks 6 through 8 calculate the area of the resolution cell, $A_{res}$, and the minimum resolution cell dimension, $\delta_{min}$. They also determine the number of ship-associated pixels (for a ship of specified dimensions) in a SAR image, $N_{ps}$, the number of pixels in the TDW, $N_{pw}$, and the maximum number of vessel-associated pixels, $N_{ps}^w$ (for a single vessel) that may be present in the TDW.

Providing a brief discussion before moving on to block 9, where the parameter $m$ is calculated, is helpful; this parameter represents the minimum number of survived pixels that must be present in a TDW to declare the ship's detection.

Using the relationship between the cumulative distribution function of the binomial distribution and the overall probability of detection $P_D$ (see, e.g., [5, p. 1033]), it is straightforward to show that for given $P_d$ and $n$, setting $m = 1$ (1-of-$n$ rule) maximizes the $P_D$ value in (17). Moreover, as demonstrated in [4, p. 339], for the 1-of-$n$ rule, the ratio $P_D/P_d \geq 1$. For $n = 1$, the ratio $P_D/P_d = 1$, and even for relatively small $n$ values, it essentially exceeds 1. For example, the 1-of-10 rule ensures the overall probability of detection, $P_D$= 0.99, for the single-trial probability of detection $P_d$= 0.369. Therefore, the ratio $P_D/P_d$ = 0.99/0.369 = 2.683. In other words, the 1-of-$n$ rule enhances the overall probability in relation to the single-trial $P_d$.

The issue with the 1-of-$n$ rule is that it has the same impact on the probability of false alarm. Indeed, as is shown in [4, p.339], for $P_{fa} \ll 1$, $P_{FA} \approx nP_{fa}$; the 1-of-$n$ rule increases the single-trial $P_{fa}$ by a factor of $n$. Therefore, it is important to find a value for the parameter $m$ that maximizes the overall probability of target detection under the constraint $P_{FA} < P_{fa}$, which indicates that the overall probability of false alarm at the target level must be less than the single-trial probability of false alarm at the pixel level.

Block 9 identifies the optimal parameter $m$ that maximizes the ship detection probability $P_D^{sw}$ while complying with the specified constraint on the false alarm probability $P_{FA}^{sw}$ at the ship level. The optimal value of $m$ defined as the smallest integer $k \in \mathcal{M} = \{2, 3, ..., N_{ps}^w\}$ that meets the following condition: $P_{FAk}^{sw} = I_{P_{fa}}(k, N_{ps}^w - k + 1) < P_{fa}$.

Block 10 accepts $\alpha'$, $\beta$, $m$, and $N_{ps}^w$ to compute the single-trial probability of detection $P_d$ using numerical integration (16). Block 11 calculates the overall probability of ship detection $P_{Dm}^{sw}$ and, finally, Block 12 stores the computed probabilities $P_{Dm}^{sw}$, $P_{FAm}^{sw}$, $P_d$, $P_{fa}$, along with parameters $m$, $\bar{\Sigma}_{sp}^0$, $\bar{X}_{sp}$, $\alpha'$, $\beta$, and other items listed in Fig. 3.

V. SHIP DETECTION FEASIBILITY ANALYSIS

This section presents an example of a feasibility study for ship detection, a critical aspect of preliminary SAR system design. This study evaluates the minimum detectable RCS, $RCS_{min}^{ship}$, for specific SAR systems and vessel parameters.

To estimate $RCS_{min}^{ship}$, we first evaluate the minimum ship-pixel associated backscattering coefficient $\min\bar{\Sigma}_{sp}^0$ and then calculate the minimum detectable RCS at the ship level as

$$RCS_{min}^{ship} = \min\bar{\Sigma}_{sp}^0 \, A_{ship} \tag{19}$$

where $A_{ship} = L_{ship} W_{ship}$ is the upper bound for the ship's area.

As can be seen, to estimate the $RCS_{min}^{ship}$, in addition to $\min\bar{\Sigma}_{sp}^0$, it is necessary to specify $\delta_{az}$, $\delta_r$, and $\psi$. To simplify the feasibility analysis, we fix the parameters $\delta_{az}$ and $\psi$, so that the $RCS_{min}^{ship}$ value in (19) depends only on $\min\bar{\Sigma}_{sp}^0$ and $\delta_r$, with other conditions being equal.

To determine the $\min\bar{\Sigma}_{sp}^0$, our method solves the following optimization problem with constraints

$$\min\bar{\Sigma}_{sp}^0 = \underset{\bar{\Sigma}_{sp}^0}{\arg\min}\left(\left|P_D - P_{Dm}^{sw}(\bar{\Sigma}_{sp}^0, \delta_r)\right|/P_D\right) \tag{20}$$
$$\bar{\Sigma}_{sp1}^0 \leq \bar{\Sigma}_{sp}^0 \leq \bar{\Sigma}_{sp2}^0, \, \delta_r \in \mathbb{S}\{\delta_r\}$$

The solution is sought within the interval $[\bar{\Sigma}_{sp1}^0, \bar{\Sigma}_{sp2}^0]$ for each fixed $\delta_r$ value in the set $\mathbb{S}\{\delta_r\}$. The maximum overall probability $P_{Dm}^{sw}(\bar{\Sigma}_{sp}^0, \delta_r)$ is calculated as shown in Fig. 3.

TABLE I
SURVEILLANCE GEOMETRY FOR VLEO SAR

| Parameter | Symbol | Value | Units |
| --- | --- | --- | --- |
| Satellite Altitude | $H_s$ | 350 | km |
| Look Angle | $LA$ | 20 | deg |
| Swath | $S$ | 20 | km |
| Max slant range | $R_{smax}$ | 377.558 | km |
| Max ground range | $R_{gmax}$ | 137.865 | km |
| Min slant range | $R_{smin}$ | 370.344 | km |
| Min ground range | $R_{gmin}$ | 117.865 | km |
| Max unamb. range | $R_{una}$ | 23.395 | km |
| Velocity (orbital/ground) | $V_{so}/V_{se}$ | 7,694/7,293 | m/s |

TABLE II
SAR SYSTEM SPECIFICATIONS

| Parameters | Symbol | X band | Ku band | Units |
| --- | --- | --- | --- | --- |
| Center frequency | $f_o$ | 9.65 | 14.85 | GHz |
| Peak Tx Power | $P_t$ | 1,400 | 1,400 | W |
| Duty factor | $d$ | 0.15 | 0.15 | – |
| PRF | PRF | 6.412 | 6.412 | kHz |
| LFM Pulse: | | | | |
|   Bandwidth | $B$ | 900 | 900 | MHz |
|   Slant Range Res. | $\delta_r$ | 0.25 | 0.25 | m |
|   Pulse width | $T_p$ | 23.39 | 23.39 | μs |
| Antenna: | | | | |
|   Dimensions | $L_h \times L_v$ | 3.0×0.8 | 3.0×0.8 | m×m |
|   3dB BW Azm/Elv | $\theta_{az}/\theta_{el}$ | 0.71/2.67 | 0.46/1.74 | deg |
|   Gain | $G$ | 43.36 | 47.10 | dBi |
| Azimuth Resolution | $\delta_{az}$ | 2.0 | 2.0 | m |
| Noise Figure | $F$ | 6.5 | 6.5 | dB |
| System Losses | $L_s$ | 7.0 | 7.0 | dB |
| System temperature | $T_o$ | 290 | 290 | K |
| Required Det. Perf. | $P_D/P_{fa}$ | 0.9 / $10^{-14}$ | 0.9 / $10^{-14}$ | – |

Tables I and II summarize the parameters of the surveillance geometry and a very low Earth orbit (VLEO) SAR we use to demonstrate the proposed method. This example considers a SAR with a fixed antenna size. For the ship of interest, we assume $\beta = 2$, $L_{ship} = 12$ m, $W_{ship} = 4$ m, and set $L_w = 6$ m for the TDW. The required probability of false alarm at the pixel level and the overall probability of detection are $P_{fa} = 10^{-14}$ and $P_D = 0.9$, respectively.

Fig. 4 illustrates the $RCS_{min}^{ship}$-vs-$\delta_r$ plots calculated at X and Ku bands for the target at the maximum slant range $R_{smax} = 377{,}558$ m ($\psi = 67.3451°$). The lower and upper bounds for $\bar{\Sigma}_{sp}^0$ used in (20) are $\bar{\Sigma}_{sp1}^0 = 0.01$ (-20 dB) and $\bar{\Sigma}_{sp2}^0 = 100$ (20 dB). The set $\mathbb{S}\{\delta_r\}$ is defined as the vector $[\delta_{r1}:\Delta\delta_r:\delta_{r2}]$, where $\delta_{r1} = 0.1$ m, $\Delta\delta_r = 0.0125$ m, and $\delta_{r2} = 0.5$ m. Generating one plot in Fig. 4 using MATLAB R2023b on a desktop with a 12th Gen Intel(R) Core (TM) i7-12700F 2.10 GHz, 64 GB RAM, and 64-bit Windows 11 takes approximately 3.5 s.

Fig. 4 shows that the minimum $RCS_{min}^{ship}$ values occur within a narrow range of high-resolution $\delta_r$ values, specifically 0.1 m $\leq \delta_r \leq$ 0.25 m. At higher range resolutions (i.e., smaller $\delta_r$), the number of ship-associated pixels $n$ in a TDW increases significantly, leading to $n \gg m$. In this scenario, the cumulative ship-level detection probability $P_{Dm}^{sw}$ reaches the desired value $P_D$ with a lower pixel-level detection probability $P_d$ than in lower-resolution instances (e.g., when $\delta_r$ is near 0.5 m).

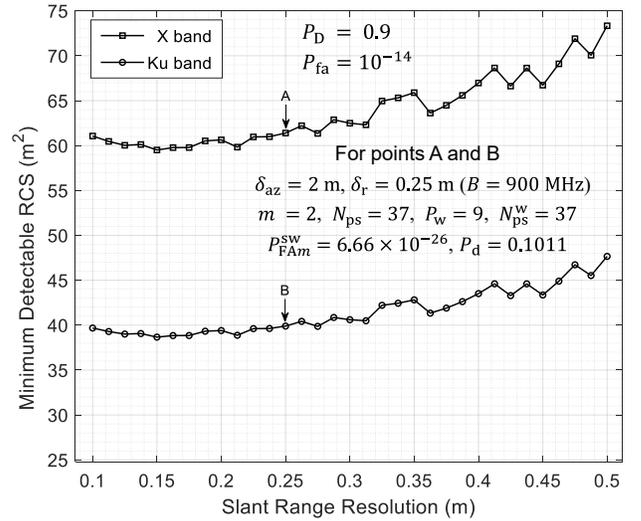

Fig. 4. Comparison of minimum detectable RCS for X and Ku band SAR systems in a VLEO satellite

A lower required $P_d$ value corresponds to a reduced SNR $\bar{X}_{sp}$ and, consequently, to a smaller value of $\min\bar{\Sigma}_{sp}^0$ and $RCS_{min}^{ship}$.

According to [1, Fig.5], the typical $\bar{\Sigma}_{sp}^0$ values range from -1 dB to 2 dB for the specified ship size. Fig. 4 indicates that for the SAR systems in question, the $\min\bar{\Sigma}_{sp}^0$ and $RCS_{min}^{ship}$ values are 1.07 dB and 61.39 m² (point A, X band), and -0.80 dB and 39.89 m² (point B, Ku band). Thus, the X band and Ku band SAR systems can detect the target of interest with the specified $P_D$ and $P_{fa}$. However, to achieve the same $\min\bar{\Sigma}_{sp}^0$ and $RCS_{min}^{ship}$ values for the X band SAR as those for the Ku band SAR, an increase in the peak power to 2,200 W is required.

Fig. 4 also shows that this example has an optimum slant range resolution $\delta_{ropt}$ near the point $\delta_r = 0.15$ m. The corresponding optimum values of $\min\bar{\Sigma}_{sp}^0$ / $RCS_{min}^{ship}$ are 0.93 dB / 59.5 m² (X band), and -0.94 dB / 38.67 m² (Ku band).

In conclusion, this letter proposes a solid criterion and an effective method for rapidly and accurately evaluating the feasibility of ship detection using spaceborne SAR systems. Generalizing this approach to any statistical model of the SAR image data and ship detection algorithms is straightforward. The proposed approach overcomes the fundamental limitations of the technique introduced in [1].